\documentclass[11pt,a4paper]{article}
\usepackage{amsmath,amssymb}
\usepackage{epsfig,graphicx}

\newcommand\ba{\begin{eqnarray}}
\newcommand\ea{\end{eqnarray}}
\newcommand\be{\begin{equation}}
\newcommand\ee{\end{equation}}
\newcommand\nn{\nonumber}
\def\thg{\theta_{\gamma}}
\newcommand{\PANDA}{$\overline{\rm P}$ANDA }
\begin{document}

\title{\bf Test of QCD through hadron form factors measurements at large momentum transfer }
\author{E.~Tomasi-Gustafsson \footnote{{\bf e-mail}: etomasi@cea.fr}
\\
\small\em  CEA, Centre de Saclay, IRFU/SPhN, 91191 Gif-sur-Yvette, France,
 and \\
\small\em Institut de Physique Nucl\'eaire, CNRS/IN2P3 and Universit\'e  Paris-Sud, France }

\date{}
\maketitle

\begin{abstract}
We review the present situation with hadron form factors in space-like and time-like regions. The possibility to transfer high momenta, and therefore to access small internal nucleon distances, allows to test nucleon models in particular pQCD predictions such as quark counting rules and helicity conservation. We will focus here to the proton case, although general statements apply also to neutron and light hadrons.

\end{abstract}

\section{Introduction}

Form factors (FFs) are fundamental quantities which describe the internal structure of composite particles. In the Nobel Lecture of R. Hofstadter, who received the Nobel Prize in 1961 for the first results obtained from electron proton ($ep$) elastic scattering measurements at SLAC, one can read: "Over a period of time lasting  at least 2000 years, 
Man has puzzled over and sought an understanding of the composition of matter". Indeed, FFs are a unique parametrization of the hadron charge and magnetic distribution. FFs are fundamental quantities, measurable through cross section and polarization phenomena. After static properties, as masses and magnetic moments, FFs constitute a good test for nucleon models as they describe dynamic properties of nuclear matter. 
The traditional way to measure electromagnetic hadron FFs is based on elastic electron proton scattering $e^-+p\to e^-+p$ and on the annihilation reactions $e^++e^-\leftrightarrow p+\bar p$. These reactions are related by the symmetries which hold for the electromagnetic and strong interactions. They are described by the same reaction amplitudes, expressed in terms of two kinematical variables, for example, the total energy $s$ and the momentum transfer squared, $q^2=-Q^2$. These variables act in different regions of the kinematical space: for annihilation reactions, in the time-like (TL) region, $q^2$ is positive and, for scattering reactions, in the space-like (SL) region, $q^2<0$.
 
Assuming that the interaction occurs through the exchange of one virtual photon, of mass $q^2$, a simple formalism relates FFs to the cross section and to the polarization observables. Assuming P and T invariance, a particle with spin $S$ is characterized by $2S+1$ FFs. Proton and neutron have two FFs, which are $a~priori$ different, a deuteron (spin one particle) has three FFs, which enter in the cross section via two structure functions. In order to measure the three deuteron FFs, it is necessary to access at least one polarization observable, usually $t_{20}$, the tensor polarization of the scattered deuteron in unpolarized $ed$ scattering. The structure of the hadronic current is similar for all hadrons, as it is driven by the reaction mechanism (one photon exchange). 

In recent years, very surprising results have been obtained in $ep$ elastic scattering, due to the possibility to apply the polarization method suggested in the sixties by A. I. Akhiezer and M. P. Rekalo \cite{Re68}: the electric and magnetic distributions in the proton have not the same dipole dependence, as a function of $q^2$ \cite{Pu10}, as it was previously assumed.

In the SL region, high precision measurements at the largest achievable values of momentum transfer squared are an important part of the experimental program at Jefferson Laboratory (USA). In the time-like region, a program is foreseen by the PANDA collaboration at FAIR (Germany), using high intensity antiproton beams up to 15 GeV momentum. Similar studies are also discussed as part of the experimental program at electron positron colliders, in Frascati (Italy), Novosibirsk (Russia), Beijing (China).

In this contribution, after a summary of recent data, the perspectives opened by such measurements at large momentum transfer will be discussed.

\section{Results and perspectives for $ep$ scattering}
In the space-like region, FFs are measured through the scattering of  polarized and unpolarized electron beams on a proton or a light nuclear target, reaching in the recent years higher precision and/or larger values of the momentum transfer. 

\subsection{Unpolarized measurements}
In case of unpolarized scattering, FFs are extracted from Rosenbluth separation, through measurements at different angles, for the same value of the momentum transfer squared (i.e., changing the initial energy and the spectrometer settings). The reduced differential cross section, after omitting known kinematical factors,  has a linear dependence as a function of $\epsilon$ where the slope is related to the electric FF,  $G_E^2$ and the intercept to the magnetic form factor, $G_M^2$:
\begin{equation}
\sigma_{red}^{Born}(\theta,Q^2)=\epsilon(1+\tau)\left [1+2\displaystyle\frac{E}{M}\sin^2(\theta/2)\right ]\displaystyle\frac
{4 E^2\sin^4(\theta/2)}{\alpha^2\cos^2(\theta/2)}\displaystyle\frac{d\sigma}{d\Omega}=
\tau G_M^2(Q^2)+\epsilon G_E^2(Q^2),
\label{eq:sigma}
\end{equation}
$$
\epsilon=[1+2(1+\tau)\tan^2(\theta/2)]^{-1}, 
$$
with the notations: 
$\alpha=e^2/(4\pi)=1/137$, $\tau=Q^2/(4M^2)>0$, $Q^2$ is the momentum transfer squared, $M$ is the proton mass, $E$ and $\theta$ are the incident electron energy and the scattering angle of the outgoing electron, respectively. 

Measurements based on this method were done up to $Q^2$=8.8 (GeV/c)$^2$ \cite{An94}. The magnetic FF was then extracted up to $Q^2\simeq 31$ (GeV/c)$^2$ \cite{Ar86}, under the definite assumption $G_E=0$. Based on these data, FFs were parametrized as 
  $G_E(Q^2)=G_M(Q^2)/\mu_p=G_D(Q^2)=(1+Q^2/m_D^2)^{-2}$, where $Q^2$ is in (GeV/c)$^2$ units, $m_D=0.71$ (GeV/c)$^2$, and $\mu=2.79$ is the proton anomalous magnetic moment in units of Bohr magneton. 
  
Common assumption was that the magnetic as well as the electric FF (although affected by larger errors) can be described by a similar dipole dependence on $Q^2$. In non relativistic approximation (and also relativistic, but in the Breit frame only) FFs are Fourier transform of the charge and magnetic distributions. Therefore, this statement is consistent with  an exponential distribution of the nucleon charge: $\rho=\rho_0e^{-r/r_0}$ with $r_0^2=(0.24~ fm)^2$ which corresponds to a mean squared radius $<r^2>\sim(0.81~fm)^2$. 
\begin{figure}
\begin{center}
\includegraphics[width=10cm]{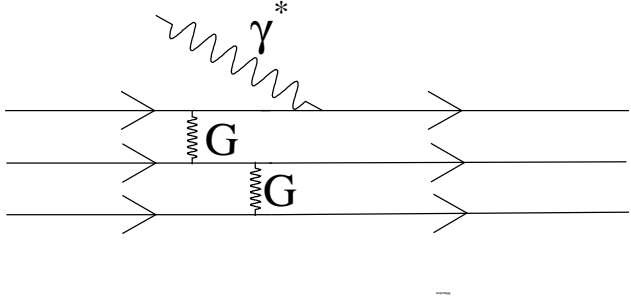}
\vspace*{.2 truecm}
\caption{pQCD diagram for $\gamma^* p$ interaction. }
 \label{Fig:pqcd}
 \end{center}
\end{figure}
On the other side, perturbative QuantumChromoDynamics (pQCD) gives a similar prediction on the $Q^2$ dependence of FFs. Assuming that the momentum is fully transfered from the incident electron through a virtual photon and that it is transmitted to each of the constituent quarks through gluon exchange, scaling laws were established (Fig. \ref{Fig:pqcd}) \cite{Ma73}.The probability to find the hadron in its ground state after that all the partons are involved in the elastic interaction, if helicity is conserved, is
$$F_n(Q^2)=C_n/[(1+Q^2/m_n)^{n-1}],$$
where $m_n=n\beta^2$, $n$ is the number of constituent quarks, and $\beta$ is the average quark momentum squared. Fitting pion data, one finds $\beta^2=0.471\pm 0.010$ (GeV/c)$^2$, which gives for the pion: 
$F_{\pi}(Q^2)=C_n/(1+Q^2/0.47)$. Rescaling by the number of quarks, 
one finds for the nucleon $F_N(Q^2)=C_n/[(1+Q^2/0.71)]^2$, which is fully consistent with the dipole parametrization, and for the deuteron $F_D(Q^2)=C_n/[(1+Q^2/1.41)^5]$,
which reproduce the existing elastic scattering data. Note that pQCD can not predict the value of $C_n$, i.e., the absolute value, but only the $Q^2$ behavior at large momentum transfer. Another strong prediction by pQCD, which is better tested through polarization measurements is that helicity should be conserved: the quark spin-flip is suppressed by an additional factor $Q^2$. The data on elastic and inelastic scattering, in the GeV range, are quite controversial on this point.

Summarizing, the dipole approximation for nucleon FFs is consistent with the unpolarized cross section measurement, and with a classical as well as a quantum view of the nucleon structure. Moreover, recent and precise measurements on unpolarized $ep$ scattering  \cite{Qa04} and re-analysis of previous data \cite{Ch04} confirm such behavior.

\begin{figure}
\begin{center}
\includegraphics[width=9cm]{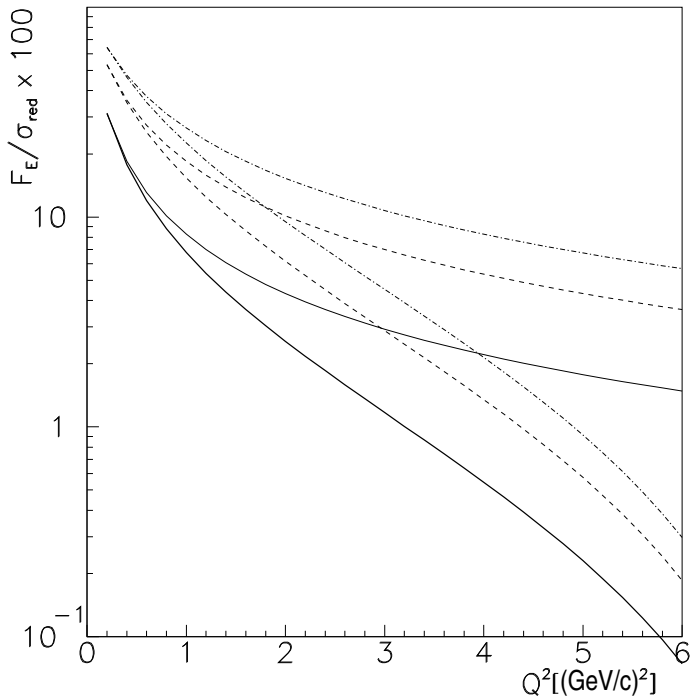}
\caption{\label{Fig:percent} Contribution of the $G_E(Q^2)$ dependent term to the reduced cross section (in percent) for $\epsilon=0.2$ (solid line), $\epsilon=0.5$ (dashed line), $\epsilon=0.8$ (dash-dotted line), in the hypothesis of FF scaling (thin lines) or following the trend suggested by polarization measurements (thick lines).}
\end{center}
\end{figure}

Note that the electric FF is affected by large error bars at large $Q^2$. This is due to the fact that magnetic part is amplified by the kinematical factor $\tau $.  Moreover, the normalization of FFs is taken to fit the static values, i.e., the charge and the magnetic moment 
$G_E(Q^2=0)=1$, $G_M(Q^2=0)=\mu_p$. This is illustrated in Fig. \ref{Fig:percent}, where the ratio of the electric part, $F_E=\epsilon G_E^2(Q^2)$, to the reduced cross section is shown as a function of $Q^2$. The different curves correspond to different values of $\epsilon$, assuming FFs scaling (thin lines).

\subsection{Polarized measurements}
High precision data on the ratio of the electric to magnetic proton FFs at large $Q^2$ have been recently obtained  \cite{Pu10}, through the polarization transfer method. In Ref.  \cite{Re68} it was firstly shown that the polarized part of the cross section, taking a longitudinally polarized electron beam and measuring the polarization of the outgoing proton in the scattering plane, contains an interference term which is proportional to the product $G_E G_M$. It is therefore much more sensitive to a small electric contribution, allowing also to determine its sign. This suggestion was far ahead the experimental possibilities. Only in the years 2000 it has been possible to achieve the necessary conditions of beam intensity, stability and polarization as well as the polarimetry technique which allows to measure the polarization of recoil protons in the GeV momentum range.
More explicitly, the longitudinal and transverse polarizations of the recoil proton are written as \cite{Re68}:
\ba
DP_t&=&-2\lambda \cot \displaystyle\frac{\theta}{2} \sqrt{\displaystyle\frac{\tau}{1+\tau }}G_EG_M,~
DP_{\ell}=\lambda\displaystyle\frac{E+E'}{M}\sqrt{\displaystyle\frac{\tau}{1+\tau }}G_M^2,~\nn \\
D&=&2\tau G_M^2+\cot^2 \displaystyle\frac{\theta}{2} \displaystyle\frac{G_E^2+\tau G_M^2}{1+\tau },
\label{eq:fi}
\ea
where $\lambda$ is the beam polarization, $E'$ is the scattered electron energy and $D$ is proportional to the differential cross section with unpolarized particles. 

The JLab $G_{Ep}$ collaboration measured the ratio $P_L/P_T$ which is directly connected to the FFs ratio:
\begin{equation}
\displaystyle\frac{P_t}{P_\ell}= - 2\cot \displaystyle\frac{\theta}{2} \displaystyle\frac{M}{E+E'}\displaystyle\frac{G_E}{G_M}.
\label{eq:final}
\end{equation}
In this ratio, the electron beam polarization as well as the analyzing powers of the polarimeter cancel, reducing the systematic errors. The same information can be derived with a polarized beam and polarized target, but the range in $Q^2$ is restricted by the limited luminosity allowed by a polarized target.

This method allows to achieve a very high precision. Moreover, the FFs data show an unexpected trend, different from what previously known from the Rosenbluth data. In Fig. \ref{Fig:G_{Ep}} (taken from Ref. \cite{Pu10}) the present status on the proton FFs is shown. The ratio $\mu G_E/G_M$ ($F_2/F_1$) from polarization experiments is shown on the top (bottom) of the figure compared with Rosenbluth measurements (green symbols). 

\begin{figure}
\begin{center}
\includegraphics[width=16 cm]{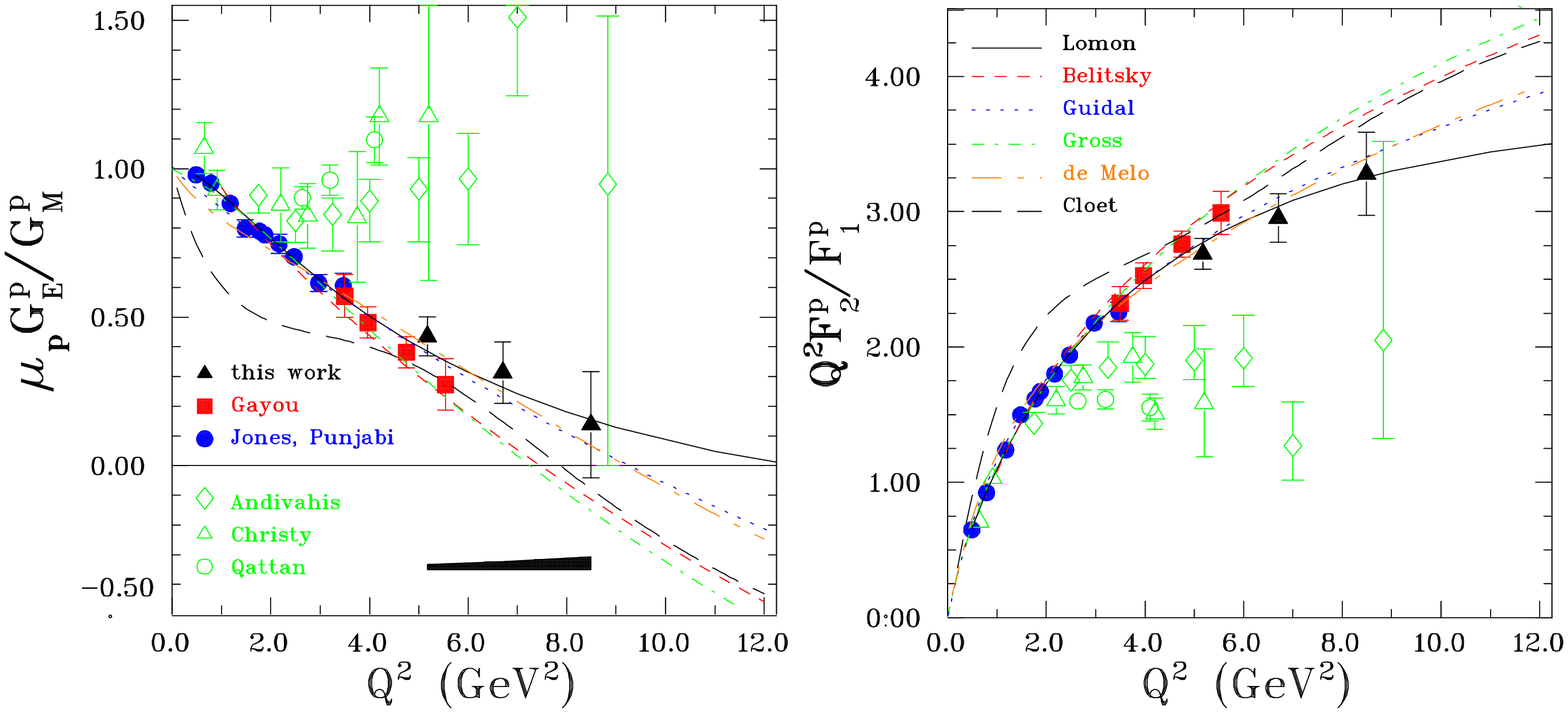}
\vspace*{.2 truecm}
\caption{ Data on proton form factor ratio as functions of $Q^2$ (from Ref. \protect\cite{Pu10}). The green symbols are from Rosenbluth experiments.}
 \label{Fig:G_{Ep}}
 \end{center}
\end{figure}
As mentioned above, PQCD predicts the asymptotic behavior $F_1\sim Q^{-4}$ and $F_2\sim Q^{-6}$ (as it involves a spin flip), which is consistent with the Rosenbluth measurements. The polarization data suggest instead the following dependence: $F_2/F_1\sim Q^{-1}$. Nucleon models, recently updated, based on very different pictures of the proton structure, are also illustrated on the figure and discussed in Ref. \cite{Pu10}.

The data obtained from the polarization method are clearly not compatible with the dipole approximation, and the difference of the ratio from unity is attributed to the electric contribution, as the magnetic FF is assumed to be well known from the cross section. In this case, the electric contribution to the cross section, illustrated in Fig. 2 (thick lines), is lower than 1\% starting at $Q^2$=3 (GeV/c)$^2$ (for $\epsilon=0.2$).

\subsection{Discussion }

The polarization data on electromagnetic FFs open important issues on the nucleon structure and arise questions on the validity of the nucleon models.
Large number of experimental and theoretical papers appeared in the recent years, on this subject.

Let us mention that not only the proton is under discussion, but also the light nuclei description. The models describing the light nuclei structure usually
assume a dipole behavior for $G_{Ep}$ and a vanishing or negligible neutron electric form factor $G_{En}$. Elastic electron-deuteron scattering is sensitive to the isoscalar form factor, $G_{Es} = G_{Ep}+G_{En}$. If $G_{Ep}$ turns out to be smaller than previously assumed, this has to be
compensated either by $G_{En}$ or by other ingredients used in deuteron calculations (which are parametrized or adjusted on the data) \cite{ETG01n}.

The discrepancy between the data extracted from these two methods is likely to be attributed to radiative corrections. The probability to irradiate one or more photons from a few GeV electron (initial, and final) may reach 40$\%$. 

Moreover, the kinematical conditions of recent experiments are very different, compared to lower intensity and/or momentum transfer. In order to compensate for the fall of the elastic rate at large $Q^2$, events are collected within wide solid angles, which implies large acceptance detectors. Not both scattered particles are momentum analyzed, in the last $G_{Ep}$ measurements the electron was detected in a calorimeter. Radiative corrections are usually applied at the lowest order in a Monte Carlo procedure, together with acceptance, efficiency and background subtraction. It is difficult to disentangle and evaluate the single effects. 

It has been recently shown that radiative corrections (RC) modify not only the yield, but also the dependence of the observables on the relevant kinematical variables. In particular, for $ep$ scattering, RC are responsible of a change of the slope (and even of its sign) of the reduced cross section \cite{ETG05b}. Let us remind that is is precisely this slope which is related to the electric FF. It has been shown that the effect of radiative corrections on the polarization ratio is small at moderate $Q^2$ \cite{Af00,By07}. 

Two photon exchange (TPE) is in principle suppressed by a factor of $\alpha$. First order corrections \cite{Mo69,Ma00} take into account TPE, when one of the photons carries all the transferred momentum, in order to cancel infrared divergences arising from soft photon emission. Long ago, it was suggested that another mechanism, where the transferred momentum is equally shared between the two photons, could enhance TPE and compensate the suppression due to the extra factor of $\alpha$ \cite{twof}. Let us stress that the inclusion of two photon effects has important consequences. The formalism which relates the cross sections to the reaction amplitudes in $ep$ scattering becomes very complicated, and the extraction of the hadron properties from the electromagnetic processes has to be revised.

Different calculations of TPE, for $ep$ elastic scattering,  give quantitatively different results according to the models, partly solving the discrepancy \cite{twofn}. Model independent properties of TPE have been derived in Ref. \cite{Re04}. The presence of TPE induces a more complicated spin structure of the matrix amplitude. In the scattering channel, instead of two real FFs, functions of one kinematical variable, $q^2$, one has to determine three amplitudes, complex functions of two kinematical variables, and the $\epsilon$ linearity of the Rosenbluth formula does not hold anymore. It has been shown that the extraction of the real FFs, is still possible but more complicate: either through a generalization of the polarization method, using electron and positron scattering on the proton in the same kinematical conditions, or measuring three T-odd or five T-even polarization observables. This is very challenging, as these observables are expected to be very small, of the order of a few percent.

In this respect, TL measurements also bring interesting and complementary information, as discussed below.

\section{Results and perspectives for $\bar p+ p \leftrightarrow e^++e^-$ scattering}
The annihilation  processes 
\be 
\bar p + p\leftrightarrow e^++ e^-
\label{eq:eq1}
\ee 
allow to access the time-like region, over the kinematical threshold, $q^2>4M^2$. The differential cross section for $ \bar p + p\to e^++ e^-$, was  first obtained in Ref. \cite{Zi62}. All polarization observables were derived in \cite{Bi93,ETG05} in frame of one photon exchange, and for TPE in Refs. \cite{Ga05,Ga06}:
\be
\displaystyle\frac{d\sigma}{d(cos\theta)} = \displaystyle\frac{\pi\alpha^2}{8M^2\tau\sqrt{\tau(\tau-1)}}
\left [ \tau |G_M|^2(1+\cos^2\theta)+ \right .
\left . |G_E|^2\sin^2\theta
\right ], ~\tau=\frac{q^2}{4M^2}>0.
\label{eq:eq5a}
\ee
Eq. (\ref{eq:eq5a}) can be simply rewritten as a linear function of $\cos^2\theta$:
\be
\displaystyle\frac{d\sigma}{d(\cos\theta)}=
\sigma_0\left [ 1+{\cal A} \cos^2\theta \right ],
\ee
in terms of the  angular asymmetry ${\cal A}$:
\be 
{\cal A}=\displaystyle\frac{\tau|G_M|^2-|G_E|^2}{\tau|G_M|^2+|G_E|^2}=
\displaystyle\frac{\tau-{\cal R}^2}{\tau+{\cal R}^2},~{\cal{R}} = \frac{|G_{E}|}{|G_{M}|}.
\label{eq:eqa}
\ee
where $\sigma_0$ is the differential cross section at 90$^\circ$:
\be
\sigma_0=\frac{\pi\alpha^2}{2q^2}\sqrt{\frac{\tau }
{\tau -1}}
\left (|G_M|^2+ \frac{1}{\tau}|G_E|^2\right ).
\label{eq:eqs0}
\ee
The linear dependence in $\cos^2\theta$ of Eq. (\ref{eq:eq5a}) results directly from the assumption of one-photon exchange, where the spin of the photon is equal to one and the electromagnetic hadron interaction satisfies $C$ invariance. Any deviation from linearity can be attributed to contributions beyond the Born approximation. In the annihilation channel, the contribution of the one--photon--exchange diagram to the reaction amplitude leads to an even function of $\cos\theta $, whereas the TPE contribution leads to four new terms reduced by an order of $\alpha $ with respect to the dominant contribution. At the reaction threshold where $q^2 = 4M^2$, one has $G_M=G_E$ and the differential cross section becomes independent on $\theta$ in the Born approximation. This is not anymore true in the presence of TPE terms. TPE  contributions are odd functions of $\cos\theta $. Therefore, they do not contribute to the differential cross section for $\theta= 90^0.$

The individual determination of the FFs in time-like region has not yet been done. From the total cross section, it is possible to extract $|G_M|$ under a definite hypothesis on $G_E$.  The 
experimental results are usually given 
in terms of a generalized FF, under the hypothesis that $G_E=0$ or $|G_E|=|G_M|$. The first hypothesis is 
arbitrary whereas the second one is strictly valid at threshold only, and there is no 
theoretical argument which justifies its validity at any other momentum 
transfer, where $q^2\neq 4M^{2}$. However, similarly to the SL region, $G_E$ plays a minor role in the cross section. Different hypothesis for  
$|G_E|$ do not affect strongly the extracted values of $G_M$, due to 
the kinematical factor 
$\tau$, which weights the magnetic contribution to the differential cross 
section and makes the contribution of the electric FF to the cross section smaller and smaller as $q^2$ increases. 

The ratio $R=G_E/G_M$ has been determined from a two parameter fit of the differential cross section, by PS170 at LEAR \cite{Ba94}, and more recently  by the BABAR Collaboration using initial state radiation (ISR), $e^++e^-\to \overline{p}+p+\gamma$ \cite{Babar}. Data are affected by large errors, mainly due to statistics. The results from BABAR suggest a ratio larger than one, in a wide region above threshold, whereas data from \cite{Ba94} suggest smaller values. 

\subsection{Perspectives at Panda-FAIR}

At the future complex accelerator FAIR, in Darmstadt, the \PANDA collaboration \cite{Panda} plans to access TL FFs through the annihilation reaction $ \bar p + p\to e^++ e^-$, 
using an antiproton beam of momentum up to 15 GeV and luminosity ${\cal L}=2\cdot 10^{32} cm^{-2}s^{-1}$.
The \PANDA detector is a ~4$\pi$ fixed target detector, designed to achieve momentum resolution at percent level for charged particles, high rate capability up to 10 MHz and good vertex resolution (~100 $\mu$m). 

\begin{figure}
\begin{center}
\includegraphics[width=10cm]{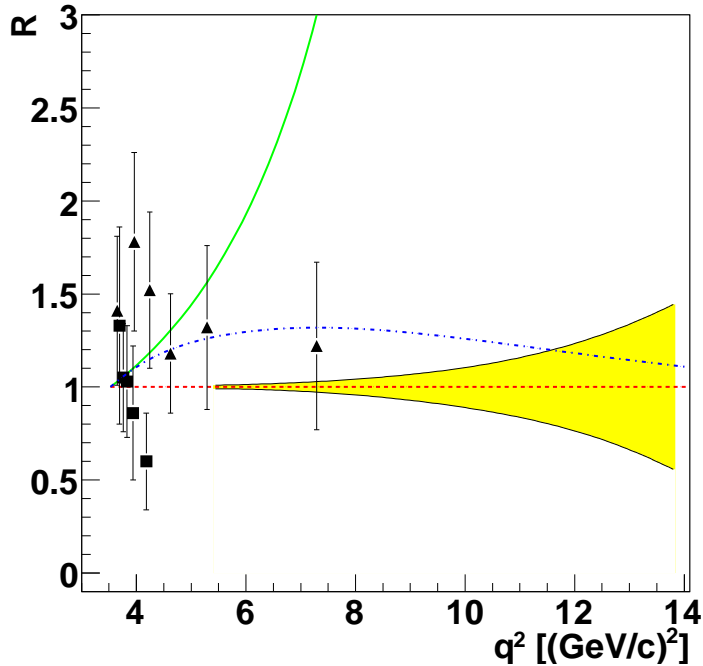}
\caption{Expected precision on the determination of the ratio ${\cal R}$, (yellow dashed band) as a function of $q^2$, compared with the existing data and to theoretical predictions. }
\label{Fig:R}
\end{center}
\end{figure}

Realistic simulations, including acceptance and efficiency, were done in a range of $q^2$ up to ~25 (GeV/c)$^2$ \cite{Su09}. The counting rates were evaluated on the basis of a realistic parametrization of FFs, assuming four months of data taking at the nominal luminosity. For each $q^2$ value, the ratio ${\cal R}=G_E/G_M$ was determined from a two parameter fit to the reconstructed spectra for the differential cross section issued from the simulation procedure.  The results are shown in Fig. \ref{Fig:R}, where the expected uncertainty on ${\cal R}$ is plotted as a function of $q^2$ as a yellow band for the case ${\cal R}=1$, to be compared with the existing values from Refs. \cite{Ba94} (squares) and \cite{Babar} (triangles). In the low $q^2$ region, the expected precision at \PANDA is at least one order of magnitude better than for the existing data and a meaningful value for ${\cal R}$ can be extracted up to at least $q^2\sim 14$ (GeV/c)$^2$. 

Nucleon models are presently little constrained and predictions display a large dispersion, as shown in Fig. \ref{Fig:R}. As an example, three models originally built in the SL region have been analytically extended to the TL region \cite{ETG05}, readjusting the parameters in order to fit the world data in all the kinematical region ($i.e.$, in SL region, the electric and magnetic proton and neutron FFs, and in TL region, the magnetic FF of the proton and the few existing data for neutron). A QCD inspired parametrization, based on scaling laws, predicts ${\cal R}=1$, as it depends only on the number of constituent quarks (dashed line). The solid line is based on a vector meson dominance (VDM) approach, and grows up to  $q^2\sim 15$ (GeV/c)$^2$. The blue-dotted line is a prediction based on an extended VDM model which includes the proper asymptotic behavior predicted by QCD.  Although these models reproduce reasonably well the FFs world data, they give different predictions for the form factor ratio and for all polarization observables. 

With a precise knowledge of the luminosity, the absolute cross section can be measured up to $q^2\sim 28$ (GeV/c)$^2$ allowing to extract a generalized FF, $|F_P| $. Comparing with the world data, one expects an improvement of at least a factor of ten (see Fig. \ref{Fig:TLSL}).

\subsection{Perspectives at BESIII-BEPCII }

Corresponding information can be accessed by the time reversed reaction $e^+  +e^-\to \bar p + p$. At the BEPCII collider, the BESIII collaboration plans precise FFs measurements. Two aspects are peculiar at BES: the possibility to scan the threshold region, and, even more interesting, the unique opportunity to measure neutron FFs: one can access in principle the neutron-antineutron final state and investigate the near threshold region where the possible contribution of a  $N\bar N$ bound state is expected.

The neutron FFs were measured at ADONE, Frascati, by the FENICE Collaboration. FFs, extracted under the hypothesis $G_E=G_M$, are twice larger than the proton FFs in the SL region, at corresponding values of the momentum transfer. The corresponding luminosity 500 nb$^{-1}$, could be reached in 15' at BES \cite{Pacetti}.

Proton FFs have been measured at BEPC by the BESII experiment, at ten center of mass energies between 2 and 3.19 GeV.  The integrated luminosity was measured by Bhabha large angle scattering, and varied from point to point from 45 to 2349 nb$^{-1}$. The maximum number of counts was lower than 30. 

Recently BABAR successfully applied the ISR method, allowing the simultaneous measurement of hadron FFs in a large $q^2$ range, when the collider is settled to study a specific resonance. Therefore it is not necessary to do an energy scan, although at the expenses of a reduction in the luminosity. The emission of a hard photon in the initial state, $e^+ + e^- \rightarrow N+ \bar N+\gamma$ allows the measurement of the non radiative processes, $e^+ + e^- \rightarrow N+ \bar N$ over a range of $N\bar N$ energy from the threshold,
$2m_N$ to the full $e^+e^-$center of mass energy, $\sqrt{s}$. The differential cross section for the radiative process, integrated over the nucleon momenta, can be factorized in a function which depends on the photon kinematical variables multiplied by the annihilation cross section for $e^+ + e^- \rightarrow N+ \bar N$:
\be
\displaystyle\frac{d^2 \sigma_{e^+e^-\to \bar pp\gamma(w)}}{dwd\cos\thg}=
\displaystyle\frac{2w}{s} W(s,x,\thg )\sigma_{p\bar p}(w)
\label{eq:eqisr}
\ee
where the total cross section for the annihilation process is function of the $p\bar p$ system invariant mass $w$:
\be
\sigma_{p\bar p}(w)=\frac{4\pi\alpha^2\beta C}{3w^2}
\left[|G_M(w)|^2+
\frac{2M^2}{w^2}|G_E(w)|^2\right ].
\label{eq:eqspp}
\ee
$\beta=\sqrt{1-4M^2/w^2}$ is the proton velocity,
$C=y/(1-e^{-y})$, and $y=\pi\alpha M/(\beta w)$ is the Coulomb correction factor.

The function $W(s,x,\thg )$ represents the probability of photon emission. It depends on the total energy $s$, on the dimensionless variable $x=\omega/\sqrt{s}$, where $\omega$ is the photon energy and of the photon emission angle $\thg$. It was used by the BABAR collaboration at large photon angles, $\thg \gg m/\sqrt{s}$ in the following form:
\be
W^{ISR}(s,x,\thg)=\displaystyle\frac{\alpha}{\pi x}
\left ( \displaystyle\frac{2-2x+x^2}{\sin^2\thg}-\displaystyle\frac{x^2}{2}\right ).
\label{eq:eqw}
\ee
Note that at zero photon emission angle, Eq. (\ref{eq:eqw}) does not apply. It has been derived from Ref. \cite{BM} by neglecting terms proportional to $m_e^2/s$ which become important at small angles. A precise  calculation in frame of the quasi-real electron method can be found in Ref. \cite{Baier}, which holds at small angles. The probability of photon emission calculated in \cite{Ben99}, generalizing the result from \cite{BM}, gives reliable results in all the angular range, and it is consistent with Ref. \cite{Baier} at small angles. 

Based on the ISR method, at 3.77 GeV center of mass energy, and luminosity $\sim 2\cdot 10^{32}$ cm$^{-2}$ s$^{-1}$ BESIII constitutes a unique opportunity to measure neutron FFs, achieving reasonable statistics in a range of moderate $q^2$ values, over the threshold.
\section{Discussion on asymptotic }

The most general form of the hadronic current is expressed in a simple form as function of the Dirac $F_1$ and Pauli $F_2$ FFs, which are linear combinations of $G_E$ and $G_M$ : $G_{E}=F_1+\tau F_2$, $~G_{M}=F_1+F_2$. It is the most suited representation for the discussion of the asymptotic behavior of FFs.

Assuming that FFs are analytical functions, one can apply the Phragm\`en-Lindel\"of theorem in the following form \cite{Ti39}:
{\it " If $f(z) \to a$ as  $z\to\infty$ 
along a straight line, 
and $f(z) \to b$ as  $z\to\infty$ 
along another 
straight line, 
and f(z) is regular 
and bounded 
in the angle between, 
then $a=b$ and $f(z) \to a$ uniformly in the angle." }

Choosing the straight lines along the $q^2$ and the  $-q^2$ axis, the application of this theorem to FFs reads:
\begin{equation}
\lim_{q^2\to -\infty} F^{(SL)}(q^2) =\lim_{q^2\to \infty} F^{(TL)}(q^2).
\label{eq:eqph}
\end{equation}
This means that, asymptotically, FFs have the following constraints: 1) the imaginary part of FFs, in TL region, vanishes: $ Im F_i(q^2)\to 0,$ as $q^2\to \infty $; 2) the real part of FFs, in TL region, coincides with the 
corresponding value in SL region, because FFs are real functions in SL region, due to the hermiticity of the corresponding electromagnetic Hamiltonian.

In order to test these requirements (based on analyticity only), the knowledge of the differential cross section for $e^++e^-\leftrightarrow p+\bar{p}$ is not sufficient, and polarization phenomena have to be studied. In this respect, T-odd polarization observables, which are determined by $Im F_1F_2^*$, are especially interesting. The simplest of these observables is the $P_y$ component of the proton polarization in $e^++e^-\to p+\bar{p}$ that in general does not vanish, even in collisions of unpolarized leptons, or the asymmetry of leptons produced in $p+\bar{p}\to e^++e^-$, in the collision of unpolarized antiprotons with a proton target polarized normally to the reaction plane (or in the collision of antiprotons polarized normally to the reaction plane on an unpolarized proton target). These observables are especially sensitive to the different parametrizations of FFs, and suggest that the corresponding asymptotics are very far \cite{ETG01,ETG05a}. 
Unfortunately, this theorem does not allow to indicate the physical value of $q^2$, starting from which it is working at some level of precision. For this aim one needs some additional dynamical information. The assumption of the analyticity of FFs allows to connect the nucleon FFs in SL and in TL regions and to extend a parametrization of FFs available in one kinematical region to the other kinematical region:
\be
t\to -t, ~ln(-t) \to ln(t)-i\pi,~t> 0.
\label{eq:eq4}
\ee

Not all nucleon models contain the intrinsic property to give real FFs in SL region, and generate an imaginary part, applying this transformation.

Dispersion relation approaches, which are based essentially on the analytical properties of nucleon electromagnetic FFs, can be considered a powerful tool for the description of the $q^2$ behavior of FFs in the entire kinematical region. The vector meson dominance (VDM)  models, can be also extrapolated from the SL region to the TL region (see \cite{ETG05} and Refs. therein). The quark-gluon string model \cite{Ka00}  allowed firstly to find the $q^2$ dependence of the electromagnetic FFs in TL region, in a definite analytical form, which can be continued in the SL region.  
\begin{figure}
\begin{center}
\includegraphics[width=11cm]{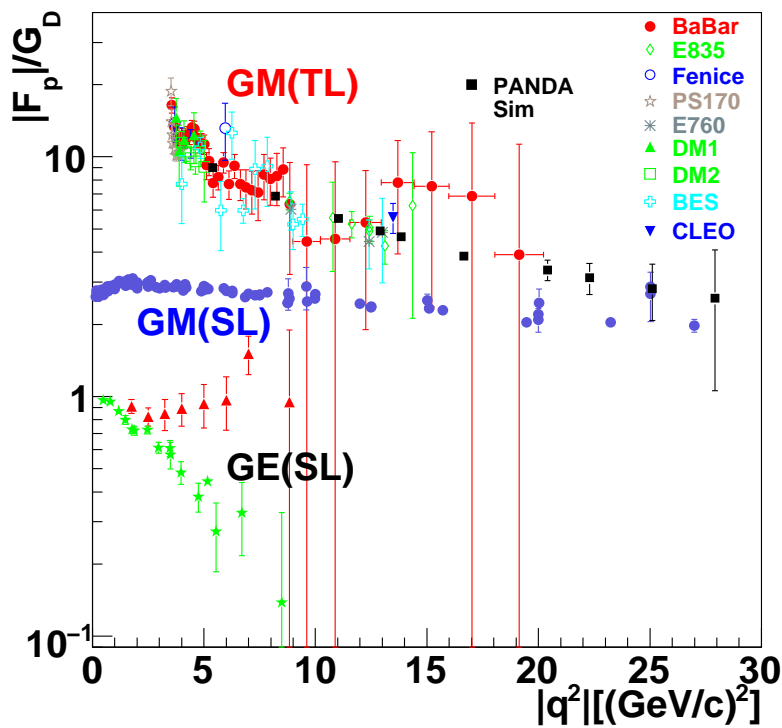}
\vspace*{.2 truecm}
\caption{World data on proton form factors, in time and space-like regions, as functions of $|q^2|$, rescaled by dipole. From top to bottom, magnetic FF in time-like region including PANDA simulated results (black solid squares), magnetic FF in space-like region (blue circles), electric FF in space-like region, from unpolarized (red triangles) and polarized (green stars) experiments.}
 \label{Fig:TLSL}
 \end{center}
\end{figure}
Experimentally it is found that the values of $G_M$ in the TL region, 
obtained under the assumption  $|G_E|=|G_M|$, are larger 
than the corresponding SL values. A difference up to a factor of two in the absolute values in SL and TL regions can be seen also for other hadron FFs, including pions and neutrons, up to the largest value at which TL FFs have been measured. 
This has been considered as a proof of the non applicability of the Phr\`agmen-Lindel\"of theorem, 
or as an evidence that the asymptotic regime is not reached \cite{Bi93}. 

A sample of the world data is reported in Fig. \ref{Fig:TLSL}, as a function of $|q^2|$, in order to overlap the SL and TL regions. From top to bottom, one can see the magnetic proton FF in TL region, under the assumption $|G_E|=|G_M|$, the magnetic proton FF in SL region which is obtained for $q^2\ge 8.8 $ (GeV/c)$^2$ under the assumption $|G_E|=0$ (blue circles) and the electric FF in SL region. Two series of data clearly show the discrepancy between unpolarized (red triangles) and polarized (green stars) measurements. 

The expected precision of the future measurements with PANDA (black solid squares) is shown in comparison with the existing data. For PANDA each point corresponds to an integrated luminosity of 2 fb$^{-1}$, which can be obtained in four months of data taking. These results have been obtained in frame of Monte Carlo simulations, which takes into account the geometry of the detector, efficiency and acceptance and it is based on a realistic parametrization of FFs  \cite{Panda}. One can see that PANDA will cover a large kinematical range and bring useful information with respect to the determination of the asymptotic region.

Note that, in principle, asymptotic properties should be discussed for $F_1$ and $F_2$, and it is not equivalent to consider, in this respect, the Sachs representation of FFs.

\section{Informative Background}

In $\bar p p$ annihilation, the selection of leptonic channels in the final state is an experimental challenge due to annihilation reactions into hadrons. Three or more hadrons in the final state can be very efficiently identified by kinematical fits. The cross section for channels involving three pions is known to be at most an order of magnitude larger than two pion production. The cross sections for the neutral (charged) channels production  are about five (six) orders of magnitude larger than the reaction of interest. In the case of the $\pi^0+\pi^0$ production, $e^-e^+$ pairs are produced  after conversion of the photons from the main $\pi^0$ decay. In addition, one (or both) $\pi^0$ may undergo Dalitz decay, $\pi^0\to\ e^-+e^++\gamma$, with probability $10^{-2}$ ($10^{-4}$). In case of charged hadron pair production, both hadrons can be misidentified as leptons. In case of kaon production, the probability of misidentification is lower and kinematical constraints are more efficient, due to their larger mass. 

On the other hand, two body hadronic reactions, which will be detected with very high statistics, contain very interesting information with respect to pQCD predictions. Following dimensional counting rules, the differential cross section of the $\bar p + p\to\pi^-+\pi^+$ process can be parametrized as \cite{Ma73,Br73}:
\be
\frac{d\sigma}{dt}=Cs^{-8}f(\theta)
\label{eq:eqsc}
\ee
where $\theta$ is the $\pi^-$ cms angle, $t$ is the Mandelstam variable and the function $f(\theta)$ depends on the reaction mechanism. In the framework of the quark interchange dominance model \cite{Gun73}, one has 
\be
f(\theta)=\frac{1}{2}(1-z^2)[2(1-z)^{-2}+(1+z)^{-2}]^{2},~z=\cos\theta.
\label{eq:eqfz}
\ee
$C=440 $ mb (GeV/c)$^{14}$ is a constant, which value is obtained from a fit to the data.

The ratio of the yields of pion and kaon pair production in the $\bar p p $ annihilation process contains interesting information on the reaction mechanism, and on the quantum numbers involved. Near the threshold region, it may allow to evidence the excitations of the physical vacuum \cite{Ku10}. Let us recall the arguments.

\begin{figure}
\begin{center}
\includegraphics[width=12cm]{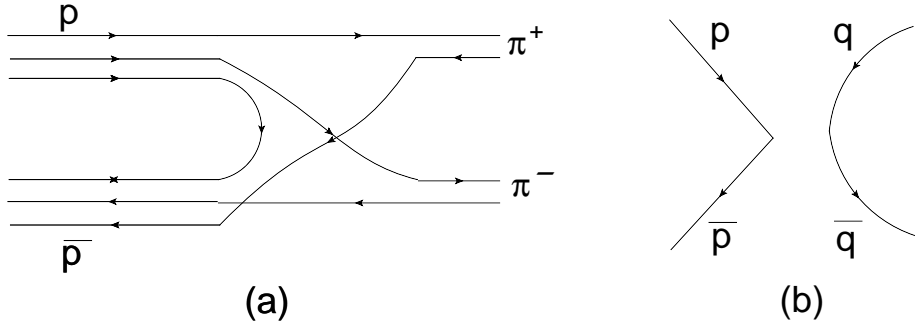}
\caption{(a) Feynman diagram for the reaction $ \bar{p} +p \to \pi^++\pi^-$ through quark rearrangement; (b) scheme for production of a pair of quarks through vacuum excitation.}
\label{Fig:ve} 
\end{center}
\end{figure}

In the threshold region, all states except $S$-states are suppressed. The main decay channel of the triplet state is to a pion charged pair, due to the absence of constituent strange quarks in the proton Fock state. On the other hand,  
the $\pi^+\pi^¯$ and $K^+K^¯$ decay channels are expected to occur with equal probability from the singlet state $^1S_0$, due to the equality of the $q\bar q$ particle-hole excitations in the physical vacuum, with $q=u,d,s$. 
Let us focus on the threshold region, neglecting the contributions from $L=1$ states. Meson production in $\bar p $ annihilation occur through rearrangement of the constituent quarks (see Fig. \ref{Fig:ve}a). Selection rules require that $J=1$, therefore, by this mechanism the $\pi^+\pi^-$ final channel occurs from $^3S_1$ state. As the strange quark content of the proton is very small, and strange quarks come only from the sea, kaon production is forbidden through such mechanism: in triplet $S$ state only pions are produced.

However a kaon pair can be produced through a disconnected diagram Fig. \ref{Fig:ve}b where any pair of particles can be created from excited vacuum. In this case, the singlet state  corresponding to $J=0$, $^1S_0$ ,can produce a pair of current quarks, which, after interacting, convert into constituent quarks and become observable as mesons. Let us underline that the probability to enhance $u\bar u$ pair of current quarks is the same as for a $s\bar s$ pair of current quarks due to the structure of excited vacuum (EV). 
The matrix element of the process $\bar p p \to \bar q q$ in singlet state, is proportional to $\bar u(p_-) v(p_+)$ (where $u,v$ are the spinors of the quark and antiquark, with four momentum $p_+$, $p_-$respectively):
\be
\left |M(\bar p p \to EV\to \bar q q)\right |^2\sim  Tr(\hat p_+-m_q)(\hat p_-+m_q)=8\beta_q^2 m_p^2,~q=u,d,s,
\ee
where $\beta_q=\sqrt{1-m^2_q/E_q^2}$,  $m_u=m_d=280$ MeV, $m_s=400$ MeV, and  $E_u=E_s=M$.

After correcting the yield of pion and kaon pair production by the ratio of phase volumes: $\phi_{\pi}/\phi_K=\beta_{\pi}/\beta_K$, one obtains
\be 
\frac{Y_{KK}}{Y_{\pi\pi}}=\frac{1/2}{3+1/2}\frac {\beta_K}{\beta_{\pi}}
\left ( \frac {\beta_s}{\beta_{u}}\right )^2=0.108.
\label{eq:eq2a}
\ee
An error of $5\%$ related to the constituent quark masses can be attributed to this value. The corresponding experimental value is $R=f(K^+K^-)/f(\pi^+\pi^-)=0.108\pm 0.007$ \cite{Ab94}. 
This estimation is based on the statement that pion and kaon pairs are produced 'democratically' from the vacuum excitation, whereas kaon production is forbidden in singlet $S$ state.

\section{Conclusions}

We have discussed future perspectives for testing QCD which are opened by the advent of high intensity accelerators and colliders with electromagnetic (electrons, positrons) and hadronic (antiprotons) probe, in relation with nucleon structure. 

In the SL region, at JLab precise experiments are ongoing/planned in order to extend FFs measurements, for proton and neutron, exploiting the qualities of the electron beam, its intensity and polarization.

The individual determination of proton FFs in a wide kinematical region will be first possible at \PANDA. Let us stress that the main advantage of this measurement in the TL region is that the dynamical information is fully contained in the angular distribution of one of the outgoing particles. This requires a single setting of the machine and of the detector (in traditional Rosenbluth measurements, at each $Q^2$ value, one has to change the energy of the beam and the angle of the outgoing particle).

The information on the reaction mechanism is also contained in the angular distribution. Due to the odd properties of the TPE contribution, TPE effects cancel (are singled out) in the sum (difference) of the cross section at complementary angles, allowing to extract the moduli of the true FFs \cite{Ga06,Ga05}. TPE effects also cancel if one does not measure the charge of the outgoing lepton. Note that evidence of TPE, based on these signatures has not been found in the experimental data in the limit of their precision: on electron elastic scattering on particles with spin zero \cite{Ga08}, one half \cite{ETG05b,Tv05}, and one \cite{Re99}. An analysis of the BABAR data \cite{Babar} does not show evidence of two photon contribution, in the limit of the uncertainty of the data \cite{ETG08}. Comparison of electron and positron elastic scattering on proton gives controversial results  \cite{ETG09,Al09,Ar04}.
New experiments at VEPP3 and DESY are on going or planned. The present simulations show that the future \PANDA experiment will be sensitive to a TPE contribution $\ge 5\%$ of the main (one photon) contribution \cite{Su09}. 

Large attention should be devoted to the problem of radiative corrections. Any reaction involving leptons is specifically affected by photon emission. This has to be carefully taken into account. High order radiative corrections play a role at large momentum transfer and may induce a specific behavior of the observables, which has to be disentangled $prior$ to extracting information on the hadron structure \cite{By07,By08}.

Polarization allows to access a larger precision and new observables. In TL region, single spin observables (in case of polarized beam or target) give information on the relative phase of FFs. The full determination of FFs requires at least the measurement of two spin observables. In case of absence of polarized beams, information on the phase of hadron FFs can be obtained for $\Lambda\bar\Lambda$ final state, which is self-polarizing through the weak decay $\Lambda\to p\pi^-$. 

\section{Acknowledgments}
The study of model independent and analytical properties of form factors was initiated by Prof. M. P. Rekalo, and based on further works with Dr. G. I. Gakh. This contribution benefits of a fruitful collaboration with Prof. E.A. Kuraev, Dr. Yu Bystritskiy, and Dr. V. Bytev. The members of the Panda group of IPN (Orsay), Prof. S. Pacetti and Prof. M. Maggiora are acknowledged for stimulating discussions on form factor measurements at Panda and BES.
The french Groupement de Recherche Nucleon, is acknowledged for useful meetings and continuous support.

{}


\begin{thebibliography}{}
\bibitem{Re68} 
  A.~I.~Akhiezer and M.~P.~Rekalo,
  Sov.\ Phys.\ Dokl.\  {\bf 13} (1968) 572
  [Dokl.\ Akad.\ Nauk Ser.\ Fiz.\  {\bf 180} (1968) 1081];  

  A.~I.~Akhiezer and M.~P.~Rekalo,
  Sov.\ J.\ Part.\ Nucl.\  {\bf 4} (1974)  277
  [Fiz.\ Elem.\ Chast.\ Atom.\ Yadra {\bf 4} (1973) 662].
\bibitem{Pu10}
  A.~J.~R.~Puckett {\it et al.},
  Phys.\ Rev.\ Lett.\  {\bf 104} (2010) 242301 and Refs therein.

\bibitem{An94} 
L.~Andivahis {\it et al.},
Phys.\ Rev.\ D {\bf 50} (1994) 5491.
\bibitem{Ar86} R. G. Arnold {\it et al.} Phys. Rev. Lett. {\bf 57}, (1986) 174. 
 \bibitem{Ma73}
V.~A.~Matveev, R.~M.~Muradian, and ~A.~N.~Tavkhelidze, Lett. Nuovo Cim.  {\bf 7} (1973) 719.

\bibitem{Qa04}
  I.~A.~Qattan {\it et al.},
  Phys.\ Rev.\ Lett.\  {\bf 94} (2005)  142301.
\bibitem{Ch04}
M.~E.~Christy {\it et al.}  [E94110 Collaboration],
Phys.\ Rev.\ C {\bf 70} (2004)  015206 .
\bibitem{ETG01n}
  E.~Tomasi-Gustafsson and M.~P.~Rekalo,
  Europhys.\ Lett.\  {\bf 55} (2001) 188.

\bibitem{ETG05b}
  E.~Tomasi-Gustafsson and G.~I.~Gakh,
  Phys.\ Rev.\  C \textbf{72} (2005) 015209.
\bibitem{Af00}
  A.~V.~Afanasev, I.~Akushevich and N.~P.~Merenkov,
  Phys.\ Rev.\ D {\bf 65} (2002)  013006.
\bibitem{By07}
  Yu.~M.~Bystritskiy, E.~A.~Kuraev and E.~Tomasi-Gustafsson,
  Phys.\ Rev.\  C {\bf 75} (2007) 015207.
\bibitem{Mo69} L. W. Mo and Y. S. Tsai, Rev. Mod. Phys. {\bf 41} (1969) 205.
\bibitem{Ma00}
  L.~C.~Maximon and J.~A.~Tjon,
  Phys.\ Rev.\ C {\bf 62} (2000) 054320 .
\bibitem{twof} 
J. Gunion and L. Stodolsky,  Phys. Rev. Lett. {\bf 30},
(1973)  345; V. Franco,  Phys. Rev. D {\bf 8}, 826 (1973); V. N. Boitsov, L.A. Kondratyuk and V.B. Kopeliovich, Sov. J.
Nucl. Phys {\bf 16} 287 (1973); F. M. Lev, Sov. J. Nucl. Phys. {\bf 21} 45 (1973).
\bibitem{twofn} 
  A.~V.~Afanasev, S.~J.~Brodsky, C.~E.~Carlson, Y.~C.~Chen and M.~Vanderhaeghen,
  Phys.\ Rev.\  D \textbf{72} (2005) 013008;
   P.~G.~Blunden, W.~Melnitchouk and J.~A.~Tjon,
   Phys.\ Rev.\  C \textbf{72} (2005) 034612;
   D.~Borisyuk and A.~Kobushkin,
   Phys.\ Rev.\  C \textbf{74} (2006) 065203.

\bibitem{Re04}
M.~P.~Rekalo and E.~Tomasi-Gustafsson,
Eur.  Phys. J. A. {\bf 22} (2004) 331;
Nucl.\ Phys.\ A {\bf 740} (2004)  271;
Nucl.\ Phys.\ A {\bf 742} 322 (2004).
\bibitem{Zi62}
A. Zichichi, S. M. Berman, N. Cabibbo, R Gatto, 
Nuovo Cim. {\bf 24} (1962) 170.
\bibitem{Bi93}
S. M. Bilenky, C. Giunti, V. Wataghin, Z. Phys. C {\bf 59} (1993) 475.
\bibitem{ETG05} 
  E.~Tomasi-Gustafsson, F.~Lacroix, C.~Duterte and G.~I.~Gakh,
Eur.\ Phys.\ J.\  A {\bf  24} (2005) 419.  
\bibitem{Ga05}
  G.~I.~Gakh and E.~Tomasi-Gustafsson,
  Nucl.\ Phys.\  A {\bf 761} (2005) 120.
\bibitem{Ga06}
  G.~I.~Gakh and E.~Tomasi-Gustafsson,
  Nucl.\ Phys.\ A \textbf{771} (2006) 169.
\bibitem{Panda} 
  Physics Performance Report for PANDA: Strong Interaction Studies with
  Antiprotons, The PANDA Collaboration,
  arXiv:0903.3905 [hep-ex]; 
  http://www.gsi.de/PANDA.  
\bibitem{Su09}
  M.~Sudol {\it et al.},
  Eur.\ Phys.\ J.\  A {\bf 44} (2010) 373.



\bibitem{Ba94}
  G.~Bardin {\it et al.},
  Nucl.\ Phys.\  B {\bf 411} (1994)  3.
\bibitem{Babar}
  B.~Aubert {\it et al.}  [BABAR Collaboration],
  Phys.\ Rev.\  D {\bf 73} (2006) 012005.
\bibitem{Pacetti} S. Pacetti, private communication.
\bibitem{BM}
  G.~Bonneau and F.~Martin,
  Nucl.\ Phys.\  B {\bf 27} (1971) 381.

 
 \bibitem{Ben99}
  M.~Benayoun, S.~I.~Eidelman, V.~N.~Ivanchenko and Z.~K.~Silagadze,
  Mod.\ Phys.\ Lett.\  A {\bf 14} (1999) 2605.


\bibitem{Baier}
  V.~N.~Baier, V.~S.~Fadin and V.~A.~Khoze,
  Nucl.\ Phys.\  B {\bf 65} (1973) 381.
  
  \bibitem{Ti39} E. C. Titchmarsh, {\it Theory of functions}, Oxford University 
Press, London, 1939, p. 179.
\bibitem{ETG01}
  E.~Tomasi-Gustafsson and M.~P.~Rekalo,
  Phys.\ Lett.\  B {\bf 504} (2001) 291.
\bibitem{ETG05a}
  E.~Tomasi-Gustafsson and G.~I.~Gakh,
  Eur.\ Phys.\ J.\  A {\bf 26} (2005) 285.
 
 \bibitem{Br73}
S.~J.~Brodsky and G.~R.~Farrar, Phys. Rev. Lett. {\bf 31} (1973) 1153.
\bibitem{Ka00}
A.~B.~Kaidalov, L.~A.~Kondratyuk and D.~V.~Tchekin,
Phys.\ Atom.\ Nucl.\  {\bf 63} (2000) 1395
[Yad.\ Fiz.\  {\bf 63} (2000)] 1474.

 \bibitem{Gun73} 
  J.~F.~Gunion, S.~J.~Brodsky and R.~Blankenbecler,
  Phys.\ Rev.\  D {\bf 8} (1973) 287.


\bibitem{Ku10}
  E.~A.~Kuraev and E.~Tomasi-Gustafsson,
  Phys.\ Rev.\  D {\bf 81} (2010) 017501.


\bibitem{Ab94}
  V.~G.~Ableev {\it et al.},
  Phys.\ Lett.\ B {\bf 329} (1994)  407.

\bibitem{Ga08}
  G.~I.~Gakh and E.~Tomasi--Gustafsson,
 Nucl.\ Phys.\  A {\bf 838} (2010) 50. 

\bibitem{Tv05}
  V.~Tvaskis, J.~Arrington, M.~E.~Christy, R.~Ent, C.~E.~Keppel, Y.~Liang and G.~Vittorini,
  Phys.\ Rev.\  C {\bf 73} (2006) 025206.
\bibitem{Re99}
  M.~P.~Rekalo, E.~Tomasi-Gustafsson and D.~Prout,
  Phys.\ Rev.\  C {\bf 60} (1999) 042202R.

\bibitem{ETG08}
  E.~Tomasi-Gustafsson, E.~A.~Kuraev, S.~Bakmaev and S.~Pacetti,
  Phys. Lett. B \textbf{659} (2008) 197.
  
\bibitem{ETG09}
  E.~Tomasi-Gustafsson, M.~O. Osipenko,  E.~A.~Kuraev, Yu.~Bystritskiy and V.~V.~Bytev,
  arXiv:0909.4736 [hep-ph] and refs. therein.

\bibitem{Al09}
  W.~M.~Alberico, S.~M.~Bilenky, C.~Giunti and K.~M.~Graczyk,
  J.\ Phys.\ G  \textbf{36} (2009) 115009.
\bibitem{Ar04}
  J.~Arrington,
  Phys.\ Rev.\  C \textbf{69} (2004) 032201.

\bibitem{By08}
  V.~V.~Bytev, E.~A.~Kuraev and E.~Tomasi-Gustafsson,
  Phys.\ Rev.\  C {\bf 77} (2008) 055205.



  
\end{thebibliography}
\end{document}